 \documentclass[]{revtex4-2}

\usepackage{pstricks,pst-all}

\usepackage{textcomp}
\usepackage{gensymb}
\usepackage{subcaption}
\usepackage{wrapfig}
\usepackage{graphicx}
\usepackage{mathtools}
\usepackage{tabularx}
\usepackage{amssymb}
\usepackage{amsmath, nccmath}
\usepackage{commath}
\usepackage{amsthm}
\usepackage{bm}
\usepackage{lineno}
\makeatletter
\newcommand*{\rom}[1]{\expandafter\@slowromancap\romannumeral #1@}
\makeatother
\usepackage{lipsum}
\makeatletter
\def\ps@pprintTitle{%
 \let\@oddhead\@empty
 \let\@evenhead\@empty
 \def\@oddfoot{}%
 \let\@evenfoot\@oddfoot}
\makeatother

\begin{document}

% \begin{frontmatter}

%% Title, authors and addresses

\title{Quasiperiodic arrangement of magnetodielectric $\delta$-plates:\\ Green's functions and Casimir energies for $N$ bodies}

%% use the tnoteref command within \title for footnotes;
%% use the tnotetext command for the associated footnote;
%% use the fnref command within \author or \address for footnotes;
%% use the fntext command for the associated footnote;
%% use the corref command within \author for corresponding author footnotes;
%% use the cortext command for the associated footnote;
%% use the ead command for the email address,
%% and the form \ead[url] for the home page:
%%
%% \title{Title\tnoteref{label1}}
%% \tnotetext[label1]{}
%% \author{Name\corref{cor1}\fnref{label2}}
%% \ead{email address}
%% \ead[url]{home page}
%% \fntext[label2]{}
%% \cortext[cor1]{}
%% \address{Address\fnref{label3}}
%% \fntext[label3]{}

%% use optional labels to link authors explicitly to addresses:
%% \author[label1,label2]{<author name>}
%% \address[label1]{<address>}
%% \address[label2]{<address>}

 \author{Venkat Abhignan}
 \address{Qdit Labs Pvt. Ltd., Bengaluru - 560092, India}
% \author{K. V. Shajesh}
% \address{School of Physics and Applied Physics,
% Southern Illinois University-Carbondale, IL - 62901, USA}
% \author{Prachi Parashar}
% \address{John A. Logan College
% Carterville, IL - 62918, USA}

\begin{abstract}
  We study a variety of finite quasiperiodic configurations with magnetodielectric $\delta$-function plates created from simple substitution rules. While previous studies for $N$ bodies involved interactions mediated by a scalar field, we extended our analysis of Green's function and corresponding Casimir energy to the electromagnetic field using plates with magnetic and dielectric properties for handling finite-size quasiperiodic lattices. The Casimir energy is computed for a class of quasiperiodic structures built from $N$ purely conducting or permeable $\delta$-plates. The Casimir energy of this quasiperiodic sequence of plates turns out to be either positive or negative, indicating that the pressure from the quantum vacuum tends to cause the stack of plates to expand or contract depending on their arrangement. We also handle the transverse electric and transverse magnetic mode Green's functions for $\delta$-plates and derive the Faddeev-like equation with the transition matrix for $N$ purely conducting or permeable plates. 
\end{abstract}

% \begin{keyword}
% Casimir interactions \sep $\delta$-function plates \sep Repulsive force
% %% keywords here, in the form: keyword \sep keyword

% %% MSC codes here, in the form: \MSC code \sep code
% %% or \MSC[2008] code \sep code (2000 is the default)

% \end{keyword}

% \end{frontmatter}

\maketitle

%%
%% Start line numbering here if you want
%%
%%\linenumbers

%% main text
\section{Introduction}

Casimir discovered that an electromagnetic force exists between two parallel dielectric plates and that changes in the vacuum energy resulting from boundary conditions of the plates on the electromagnetic field can be interpreted as the source of the associated energy \cite{Casimir:1948dh, Lifshitz:1956zz}. Even while precise Casimir force has been obtained for multiple bodies with ideal boundary conditions \cite{BALIAN1977,BALIAN1978}, it is known that the non-additivity of this interaction makes it challenging to compute Casimir force for finite dielectric properties \cite{Ninham1,Ninham2}. Casimir force for multiple planar bodies was initially pursued by Toma\ifmmode \check{s}\else \v{s}\fi{} \cite{mult3,mult4}. Further, it was observed recently that the Casimir energy of the two bodies can be concisely described using the transition matrices of the two bodies and free Green's function \cite{Kenneth2006,PRLemig}. Using this idea, the $N$ body  Casimir energy was expressed in terms of the $N$ body transition matrix for scalar field \cite{Shajesh2011,selfsimilar}. Similarly, magnetodielectric $\delta$-function plates mediated by the electromagnetic field were studied \cite{Prachi2012,Milton2013}, and the Casimir energy for $N$ plates was distributed into nearest neighbour scattering and next-to-nearest neighbour scattering terms \cite{Abhignan_2023}.

  Handling Green's function for multiple bodies \cite{mult1,mult2} helps in the calculation of the Casimir force for multiple bodies. The derivation of $N$ body Green's functions for $\delta$-plates mediated by a scalar field has led to the derivation of Faddeev-like equation \cite{Faddeev1965,faddeev1993quantum} for the transition matrix of $N$ body with transition matrices of individual bodies \cite{Shajesh2011}. These equations have also been defined recursively to obtain Green's functions for $N$ bodies \cite{selfsimilar}. Further, they developed and studied a Casimir energy formalism for a class of self-similar systems that can be applied to configurations without equally spaced plates. It is ideal for infinitely many plates with scaling behaviour in space.
  
  A fundamental inquiry in quasiperiodic systems examines the relation between the topological order and the system's structurally derived physical qualities \cite{PhysRevLett.53.1951,PhysRevLett.53.2477,10.1093/oso/9780198513896.001.0001}. The dynamics of elementary excitations, such as electrons, phonons, excitons, polaritons, spin waves, plasmons, or magnons in quasiperiodic lattices, have yielded some significant results \cite{maciá2000electrons,ALBUQUERQUE2003,albuquerque2004polaritons,Maciá_2006,Vardeny2013,10.1093/oso/9780198824442.001.0001}. Nearest-neighbour interaction models have been studied extensively in quasiperiodic systems. When concerned with quantum fields interacting with external boundaries, the Casimir effect has become an essential subject in fields like nanotechnology and biophysics \cite{RevModPhys2009,RevModPhys2016,RevModPhys2018}. The interplay between quasiperiodic order and Casimir interactions has not been explored much. Casimir interaction alters the vacuum energy, which is influenced by the space's geometry, topology, or dimension. Consequently, the vacuum energy reflects the space's underlying geometric properties and spectral characteristics. 
  
  Initially, the Casimir interaction was realized as an attractive force \cite{Casimir:1948dh}; however, later, it was discovered that the Casimir force could turn repulsive \cite{Boyer1974}, which led to a study of its practical realization, and its theoretical significance is considered essential \cite{Hoye2018,Brevik2018}. While Casimir's attractive force is recognized between two perfect electrical conductors, Boyer initially derived the repulsive force between a perfect electrical conductor and a perfect magnetic conductor \cite{Boyer1974}. Fundamentally, it was theorized that the spectral distribution of scattering responsible for attractive force is of Bose-Einstein statistics, whereas Fermi-Dirac statistics for repulsive force \cite{Brevik2018}. In a vacuum, we realize these distributions from an interaction between $\delta$-function plates \cite{Prachi2012,Milton2013} and investigate the quasiperiodic configurations with fundamentally different statistics. Utilizing previous work \cite{Abhignan_2023}, we have formulated methods to handle Green's function and Casimir energy of the quasiperiodic lattice with finite and constant dielectric or magnetic properties describing the spectral distribution of scattering. In Sec.\rom{2}, the transverse electric and transverse magnetic mode Green's functions derived previously for $N=1,2,3$ $\delta$-function plates configurations were used to obtain Faddeev-like equation with the transition matrix for $N$ plates in terms of the individual plates that are purely conducting or permeable similar to scalar case \cite{Shajesh2011}. We generalize the $N$-body Green's function for magnetodielectric $\delta$-plates. In Sec.\rom{3}, we study the Casimir energy for a class of finite quasiperiodic configurations with magnetodielectric $\delta$-plates that are equally spaced and built from purely conducting or permeable $\delta$-plates. Initially, we study Casimir energies of quasiperiodic sequences defined in terms of substitution rules \cite{Maciá_2006} for perfectly conducting or permeable $\delta$-plates (Dirichlet or Neumann boundary conditions for scalar field with only nearest-neighbor interaction) and then generalize the study to plates with finite properties (with the nearest neighbor, next-to-nearest neighbor, and all interactions). However, while previous studies involved studying scalar Casimir energy of self-similar plates exactly at infinite iterations \cite{selfsimilar}, our study with magnetodielectric properties in quasiperiodic lattices was performed only at the first few iterations numerically, which might change at the infinite limit. Our interest was limited to practical finite-size lattices with truncated self-similar behavior.

\section{Green's function and Casimir energy of magnetodielectric $\delta$-plates} 
     Second-order differential equations for electric and magnetic fields corresponding to transverse electric (TE) and transverse magnetic (TM) modes can be obtained by decoupling Maxwell's equations for $\delta$-function plates \cite{Prachi2012,Milton2013}. Green's dyadic in Fourier space of $x-y$ axis are represented as Green's functions, which correlate the electric and magnetic fields at two distinct points in space. Differential equations such as \begin{equation}
          \left[ - \frac{\partial}{\partial z} \frac{1}{\mu^\perp(z)}
\frac{\partial}{\partial z} + \frac{k_\perp^2}{\mu^{||}(z)} -\omega^2 \varepsilon^\perp(z) \right] g^E(z,z^\prime) = \delta(z-z^\prime) \label{18}
      \end{equation} and  \begin{equation}
         \left[ - \frac{\partial}{\partial z} \frac{1}{\varepsilon^\perp(z)}
\frac{\partial}{\partial z} + \frac{k_\perp^2}{\varepsilon^{||}(z)} -\omega^2 \mu^\perp(z) \right] g^H(z,z^\prime) = \delta(z-z^\prime), \label{19}
      \end{equation} with lateral wavenumbers $k_\perp$ and frequency $\omega$ are obtained by defining the electric Green's function $g^E$, which refers to the TE mode, and the magnetic Green's function $g^H$, which refers to the TM mode. Here, dielectric permittivity $\bm{\varepsilon} (z) = \varepsilon^\perp (z)\, {\bf 1}_\perp 
+ \varepsilon^{||}(z) \, \hat{\bf z} \,\hat{\bf z}$ and
magnetic permeability $\bm{\mu} (z) = \mu^\perp (z)\, {\bf 1}_\perp 
+ \mu^{||} (z) \,\hat{\bf z} \,\hat{\bf z}$ are the properties of the materials. Based on the Euclidean postulate in the frequency domain, we use $\omega\rightarrow\hbox{i}\zeta$ to rotate from Euclidean to imaginary frequencies \cite{Milton:1978sf,schwinger1988particles}.

Magnetodielectric $\delta$-plates at positions $z=a_i$ in vacuum are defined to have electric and magnetic properties
\begin{subequations}
\begin{eqnarray}
    \bm{\varepsilon}(z)=\mathbf{1}+\boldsymbol{\lambda}_{ei}(\zeta)\delta(z-a_i), \\
    \bm{\mu}(z)=\mathbf{1}+\boldsymbol{\lambda}_{gi}(\zeta)\delta(z-a_i).
\end{eqnarray}
\label{32}%%%%%%%%%%
\end{subequations}
The $\delta$-plate with planar symmetry are described by \begin{equation}
  \boldsymbol{\lambda}(\zeta) = \left[
\begin{array}{ccc}
 \lambda^{\perp}(\zeta) & 0 & 0 \\
 0 & \lambda^{\perp}(\zeta) & 0 \\
 0 & 0 & 0 \\
\end{array}
\right],
\end{equation} which implies homogeneous and isotropic properties on the $x-y$ plane. Green's functions $g^E$ (Eq. 1) and $g^H$ (Eq. 2) are obtained using boundary conditions
\begin{subequations}
\begin{eqnarray}
g^E(z,z^\prime) \Big|^{z=a_i+\delta}_{z=a_i-\delta}
&=& \frac{\lambda^\perp_{g i}}{2}
\left[ \left\{ \frac{\partial}{\partial z}
g^E(z,z^\prime) \right\}_{z=a_i+\delta} 
+ \left\{\frac{\partial}{\partial z}
g^E(z,z^\prime) \right\}_{z=a_i-\delta} \right], \\
 \frac{\partial}{\partial z} g^E(z,z^\prime)\bigg|^{z=a_i+\delta}_{z=a_i-\delta}
&=& \zeta^2 \frac{\lambda^\perp_{e i}}{2}
\big[ g^E(a_i+\delta,z^\prime) + g^E(a_i-\delta,z^\prime) \big].
\end{eqnarray}
\label{37}
\end{subequations}
and
\begin{subequations}
\begin{eqnarray}
g^H(z,z^\prime) \Big|^{z=a_i+\delta}_{z=a_i-\delta}
&=& \frac{\lambda^\perp_{e i}}{2}
\left[ \left\{  \frac{\partial}{\partial z}
g^H(z,z^\prime) \right\}_{z=a_i+\delta} 
+ \left\{  \frac{\partial}{\partial z}
g^H(z,z^\prime) \right\}_{z=a_i-\delta} \right], \\
  \frac{\partial}{\partial z}
g^H(z,z^\prime) \bigg|^{z=a_i+\delta}_{z=a_i-\delta}
&=& \zeta^2 \frac{\lambda^\perp_{g i}}{2}
\big[ g^H(a_i+\delta,z^\prime) + g^H(a_i-\delta,z^\prime) \big],
\end{eqnarray}
\label{38}%%%%%%%%%%%
\end{subequations}
respectively.
\subsection{Green's function of $N$ magnetodielectric $\delta$-plates}
\begin{figure}[!ht]
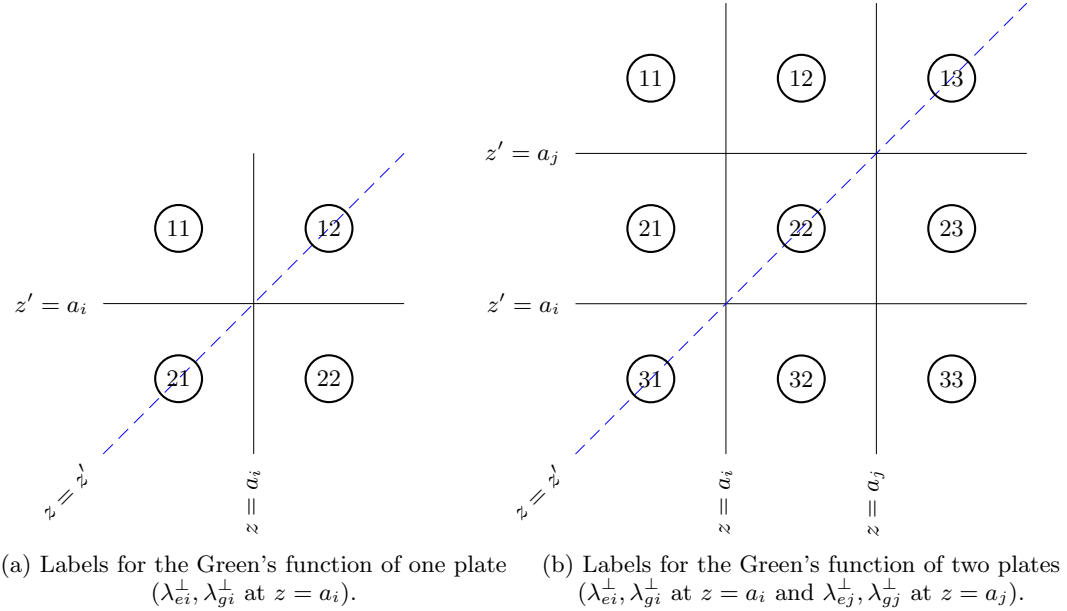

 \centering
 \begin{subfigure}{0.4\textwidth}
 \pspicture(0,0)(4,-4.5)

\psline[linecolor=black,linewidth=0mm]{-}(0,-2)(4,-2) 
\psline[linecolor=black,linewidth=0mm]{-}(2,-0)(2,-4) 
\psline[linecolor=blue,linewidth=0mm,linestyle=dashed,
strokeopacity=0.6]{-}(0,-4)(4,-0) 
\rput[r]{45}(-0.2,-4.2){$z=z^\prime$}
\rput(1,-1){\pscirclebox{11}}
\rput(1,-3){\pscirclebox{21}}
\rput(3,-3){\pscirclebox{22}}
\rput(3,-1){\pscirclebox{12}}
\rput[r]{90}(2.0,-4.2){$z=a_i$}
\rput[r]{0}(-0.2,-2.0){$z^\prime=a_i$}
\endpspicture
\vspace{6mm}
 \caption{Labels for the Green's function of one  plate ($\lambda^{\perp}_{e i}, \lambda^{\perp}_{g i}$ at $z=a_i$).}
 \label{1a}
 \end{subfigure}
 \begin{subfigure}{0.4\textwidth}
\pspicture(0,0)(6,-6.5)

\psline[linecolor=black,linewidth=0mm]{-}(0,-2)(6,-2) 
\psline[linecolor=black,linewidth=0mm]{-}(0,-4)(6,-4) 
\psline[linecolor=black,linewidth=0mm]{-}(2,-0)(2,-6) 
\psline[linecolor=black,linewidth=0mm]{-}(4,-0)(4,-6) 
\psline[linecolor=blue,linewidth=0mm,linestyle=dashed,
strokeopacity=0.6]{-}(0,-6)(6,-0) 
\rput(5,-5){\pscirclebox{33}}
\rput(1,-5){\pscirclebox{31}}
\rput(1,-1){\pscirclebox{11}}
\rput(5,-1){\pscirclebox{13}}
\rput(3,-5){\pscirclebox{32}}
\rput(1,-3){\pscirclebox{21}}
\rput(3,-3){\pscirclebox{22}}
\rput(5,-3){\pscirclebox{23}}
\rput(3,-1){\pscirclebox{12}}
\rput[r]{90}(2.0,-6.2){$z=a_i$}
\rput[r]{90}(4.0,-6.2){$z=a_j$}
\rput[r]{0}(-0.2,-2.0){$z^\prime=a_j$}
\rput[r]{0}(-0.2,-4.0){$z^\prime=a_i$}
\rput[r]{45}(-0.2,-6.2){$z=z^\prime$}
\endpspicture
\vspace{6mm}
 \caption{Labels for the Green's function of two plates ($\lambda^{\perp}_{e i}, \lambda^{\perp}_{g i}$ at $z=a_i$ and $\lambda^{\perp}_{e j}, \lambda^{\perp}_{g j}$ at $z=a_j$).}
 \label{1b}
 \end{subfigure}
 \caption{ Different labels for the regions in the $z-z'$ space for the Green's functions of thin magnetodielectric plates.}\label{Fig1}
 \end{figure}

\begin{figure}[htp]
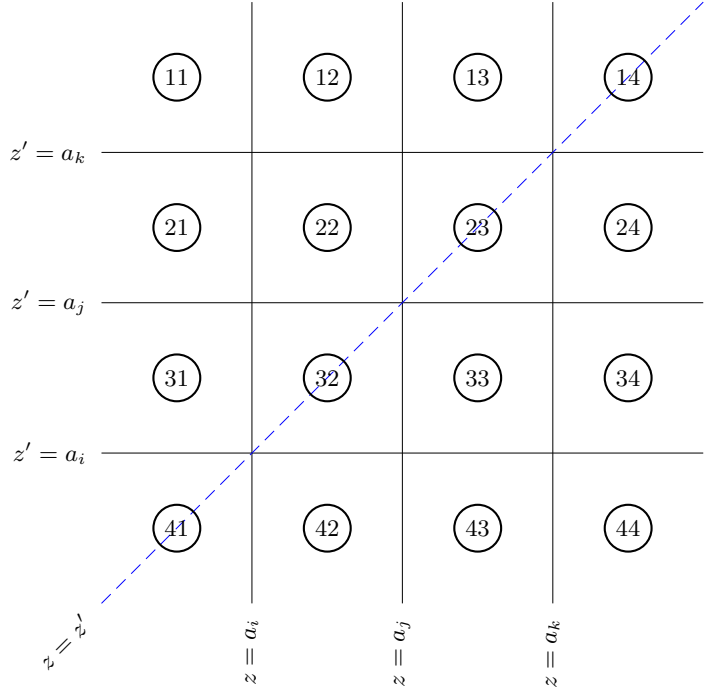

     \begin{center}
      \pspicture(0,0)(8,-8.5)
\psline[linecolor=black,linewidth=0mm]{-}(0,-2)(8,-2) 
\psline[linecolor=black,linewidth=0mm]{-}(0,-4)(8,-4) 
\psline[linecolor=black,linewidth=0mm]{-}(0,-6)(8,-6) 
\psline[linecolor=black,linewidth=0mm]{-}(2,-0)(2,-8) 
\psline[linecolor=black,linewidth=0mm]{-}(4,-0)(4,-8) 
\psline[linecolor=black,linewidth=0mm]{-}(6,-0)(6,-8) 
\psline[linecolor=blue,linewidth=0mm,linestyle=dashed,
strokeopacity=0.6]{-}(0,-8)(8,-0) 
\rput(5,-5){\pscirclebox{33}}
\rput(1,-5){\pscirclebox{31}}
\rput(1,-1){\pscirclebox{11}}
\rput(5,-1){\pscirclebox{13}}
\rput(3,-7){\pscirclebox{42}}
\rput(7,-3){\pscirclebox{24}}
\rput(1,-7){\pscirclebox{41}}
\rput(3,-5){\pscirclebox{32}}
\rput(1,-3){\pscirclebox{21}}
\rput(3,-3){\pscirclebox{22}}
\rput(5,-3){\pscirclebox{23}}
\rput(3,-1){\pscirclebox{12}}
\rput(7,-1){\pscirclebox{14}}
\rput(5,-7){\pscirclebox{43}}
\rput(7,-7){\pscirclebox{44}}
\rput(7,-5){\pscirclebox{34}}
\rput[r]{90}(2.0,-8.2){$z=a_i$}
\rput[r]{90}(4.0,-8.2){$z=a_j$}
\rput[r]{90}(6.0,-8.2){$z=a_k$}
\rput[r]{0}(-0.2,-2.0){$z^\prime=a_k$}
\rput[r]{0}(-0.2,-4.0){$z^\prime=a_j$}
\rput[r]{0}(-0.2,-6.0){$z^\prime=a_i$}
\rput[r]{45}(-0.2,-8.2){$z=z^\prime$}
\endpspicture
\vspace{6mm}
     \caption{Labels for the Green's function of three plates ($\lambda^{\perp}_{e i}, \lambda^{\perp}_{g i}$ at $z=a_i$, $\lambda^{\perp}_{e j}, \lambda^{\perp}_{g j}$ at $z=a_j$ and $\lambda^{\perp}_{e k}, \lambda^{\perp}_{g k}$ at $z=a_k$).}
    \label{Fig2}
    \end{center}
 \end{figure}

For $N=1,2,3$ $\delta$-plates, Green's function in various regions of $z-z'$ space is illustrated in Figs. (\ref{1a}), (\ref{1b}), and (\ref{Fig2}) for the configuration of $N=1$ plate with properties $\lambda^{\perp}_{e i}, \lambda^{\perp}_{g i}$ at $z=a_i$, $N=2$ plates with properties $\lambda^{\perp}_{e i}, \lambda^{\perp}_{g i}$ at $z=a_i$, $\lambda^{\perp}_{e j}, \lambda^{\perp}_{g j}$ at $z=a_j$ and $N=3$ plates with properties $\lambda^{\perp}_{e i}, \lambda^{\perp}_{g i}$ at $z=a_i$, $\lambda^{\perp}_{e j}, \lambda^{\perp}_{g j}$ at $z=a_j$, $\lambda^{\perp}_{e k}, \lambda^{\perp}_{g k}$ at $z=a_k$, respectively. The matrices $A$, $B$, and $C$ were solved in connection to these regions, which depicts the propagation of multiple paths in Green's functions \cite{Abhignan_2023}. TM mode Green's function for $N$ plate configuration $g^{N,H}$ was obtained in the form of these matrices $A$, $B^H$, and $C$ as
 \begin{equation}
	 g^{N,H}_{ \rput(0.1,0){\scriptstyle{ij}} \pscircle[linewidth=0.2pt](0.1,0.01){0.15} } =
	\begin{cases}
		\frac{1}{2 \kappa} \hbox{e}^{-\kappa|z-z'|}+\frac{1}{2 \kappa} A_{i} B^H_{ij} C_{j}& 
		\text{if}\hspace{2mm} i+j-2=N,\\
		\frac{1}{2 \kappa} A_{i} B^H_{ij} C_{j}&  \text{if}\hspace{2mm} i+j-2\neq N.
	\end{cases}
	\label{matrix}
\end{equation} $B^H_{ij}$ represents the element of a matrix $B^H$ in $i^{th}$ row, $j^{th}$ column for $i=1,\cdots,N+1$, $j=1,\cdots,N+1$ and $N$ denotes the number of plates. Upon deriving TE mode Green's function $g^E$, they have expressions that are identical to $g^H$ and can be obtained by switching $\lambda^{\perp}_{e i} \leftrightarrow \lambda^{\perp}_{g i}$ and replacing superscripts $H \rightarrow E$.

The matrices $A$, $B^H$, and $C$ representing distinct regions in $z-z'$ space for $g^{1,H}$ of $N=1$ plate configuration with the subscript label in Fig. (\ref{1a}) are
\begin{equation}
    A = \left[\begin{array}{c c}
 \hbox{e}^{-\kappa(z'-a_i)} &  \hbox{e}^{-\kappa(a_i-z')} \end{array}\right], B^H = \left[\begin{array}{c c} t_i^H & r_i^H \\ r_i^H & t_i^H
\end{array}\right]\,\hbox{and}\, C = \left[\begin{array}{c c}
\hbox{e}^{-\kappa(a_i-z)} \\ \hbox{e}^{-\kappa(z-a_i)} \end{array}\right]. \label{mat}
\end{equation}
Green's function $ g^{1,H}_{ \rput(0.1,0){\scriptstyle{ij}}
 \pscircle[linewidth=0.2pt](0.1,0.01){0.15} }$ in various regions of $z-z'$ space can be obtained here from Eq. (\ref{matrix}) for $N=1$. The reflection $r_i$ and transmission $t_i$ coefficient of plates are \begin{align}
        r_i^H= -\frac{\lambda^{\perp}_{g i} \zeta^2}{\lambda^{\perp}_{g i} \zeta^2+2 \kappa}+\frac{\lambda^{\perp}_{e i} \kappa}{\lambda^{\perp}_{e i} \kappa +2 }, \ t_i^H=1-\frac{\lambda^{\perp}_{g i} \zeta^2}{\lambda^{\perp}_{g i} \zeta^2+2 \kappa}-\frac{\lambda^{\perp}_{e i} \kappa}{\lambda^{\perp}_{e i} \kappa +2 }
        \label{coeff}
         \end{align} with $\kappa=\sqrt{k_{\perp}^2+\zeta^2}$.

Similarly the matrices $A$, $B^H$, and $C$ representing distinct regions in $z-z'$ space for $g^{2,H}$ and $g^{3,H}$ of $N=2$ and $N=3$ plates configurations in Fig. (\ref{1b}) and Fig. (\ref{Fig2}), respectively are given in the appendix A. The cumbersome expressions for the distinct  Green's functions in $(N+1)^2$ regions for $N$ plates in $z-z'$ space are better represented using these matrices. For instance, reflection coefficients for $N=1$ configuration can be obtained from  $B^H_{12}$, $B^H_{21}$ in $g^{1,H}_{\rput(0.1,0){\scriptstyle{ij}}
 \pscircle[linewidth=0.2pt](0.1,0.01){0.15}}$, whereas reflection coefficients for $N=2$ configuration can be obtained from $B^H_{13}$, $B^H_{31}$ in $g^{2,H}_{ \rput(0.1,0){\scriptstyle{ij}}\pscircle[linewidth=0.2pt](0.1,0.01){0.15} }$ and reflection coefficients for $N=3$ configuration can be obtained from  $B^H_{14}$, $B^H_{41}$ in $g^{3,H}_{\rput(0.1,0){\scriptstyle{ij}}
 \pscircle[linewidth=0.2pt](0.1,0.01){0.15}}$. Similarly, the transmission coefficients for $N=1$ configuration can be obtained from  $B^H_{11}$, $B^H_{22}$ in $g^{1,H}_{\rput(0.1,0){\scriptstyle{ij}}
 \pscircle[linewidth=0.2pt](0.1,0.01){0.15}}$, whereas transmission coefficients for $N=2$ configuration can be obtained from $B^H_{11}$, $B^H_{33}$ in $g^{2,H}_{ \rput(0.1,0){\scriptstyle{ij}}\pscircle[linewidth=0.2pt](0.1,0.01){0.15} }$ and transmission coefficients for $N=3$ configuration can be obtained from  $B^H_{11}$, $B^H_{44}$ in $g^{3,H}_{\rput(0.1,0){\scriptstyle{ij}}
 \pscircle[linewidth=0.2pt](0.1,0.01){0.15}}$. The remaining terms represent the various possibilities for the path of propagation exponentially depending on the length of propagation $|z-a_i|$ and $|z'-a_j|$ in the particular region of $g_{\rput(0.1,0){\scriptstyle{ij}}
 \pscircle[linewidth=0.2pt](0.1,0.01){0.15} }$ \ .

For understanding the physical interpretation between the regions in Fig. (1a), let us consider regions $g^{1,H}_{ \rput(0.1,0){\scriptstyle{11}}
  \pscircle[linewidth=0.2pt](0.1,0.01){0.15} }$ and $  g^{1,H}_{ \rput(0.1,0){\scriptstyle{12}}
  \pscircle[linewidth=0.2pt](0.1,0.01){0.15} }$ of one $\delta$-plate where \begin{subequations}
  \begin{align}
  g^{1,H}_{ \rput(0.1,0){\scriptstyle{11}}
  \pscircle[linewidth=0.2pt](0.1,0.01){0.15} }
  (z,z^\prime) &= \frac{t_i^H}{2 \kappa} \hbox{e}^{-\kappa(a_i-z)} \hbox{e}^{-\kappa(z'-a_i)},
  && z<a_i<z^\prime \\
  g^{1,H}_{ \rput(0.1,0){\scriptstyle{12}}
  \pscircle[linewidth=0.2pt](0.1,0.01){0.15} }
  (z,z^\prime) &= \frac{1}{2 \kappa} \hbox{e}^{-\kappa|z-z'|}+\frac{r_i^H}{2 \kappa} \hbox{e}^{-\kappa(z-a_i)} \hbox{e}^{-\kappa(z'-a_i)},
  && z,z^\prime> a_i.
  \end{align}
  \end{subequations}
Considering the point of source as $z'$ and point of observation as $z$, region $g^{1,H}_{ \rput(0.1,0){\scriptstyle{11}}
  \pscircle[linewidth=0.2pt](0.1,0.01){0.15} }$ gives the transmission amplitude of the propagator across the interface $\lambda^{\perp}_{e i}, \lambda^{\perp}_{g i}$ at $z=a_i$. The two terms in $g^{1,H}_{ \rput(0.1,0){\scriptstyle{12}}
  \pscircle[linewidth=0.2pt](0.1,0.01){0.15} }$; the first term refers to the propagator directly from the source to point of observation, and the second term represents the propagator reflected at the interface and back to the observation point with a reflection amplitude from the interface.

Similarly, to understand the physical interpretation between the regions in Fig. (1b) for two $\delta$-plates, let us consider regions
\begin{subequations}
  \begin{align}
  g^{2,H}_{ \rput(0.1,0){\scriptstyle{11}}
  \pscircle[linewidth=0.2pt](0.1,0.01){0.15} }
  (z,z^\prime) &= \frac{t_i^H \ \hbox{e}^{-\kappa a} \ t_j^H}{2 \kappa \Delta_{12}^H} \hbox{e}^{-\kappa(a_i-z)} \hbox{e}^{-\kappa(z'-a_j)},
  && z<a_i<a_j<z^\prime, \\
  g^{2,H}_{ \rput(0.1,0){\scriptstyle{12}}
  \pscircle[linewidth=0.2pt](0.1,0.01){0.15} }
  (z,z^\prime) &= \frac{t_j^H}{2 \kappa \Delta_{12}^H} \hbox{e}^{-\kappa(a_j-z)} \hbox{e}^{-\kappa(z'-a_j)}+\frac{r_i^H \ \hbox{e}^{-\kappa a} \ t_j^H}{2 \kappa \Delta_{12}^H} \hbox{e}^{-\kappa(z-a_i)} \hbox{e}^{-\kappa(z'-a_j)},
  && a_i<z<a_j<z^\prime, \\
  g^{2,H}_{ \rput(0.1,0){\scriptstyle{13}}
  \pscircle[linewidth=0.2pt](0.1,0.01){0.15} }
  (z,z^\prime) &= \frac{1}{2 \kappa} \hbox{e}^{-\kappa|z-z'|}+\frac{r_j^H}{2 \kappa} \hbox{e}^{-\kappa(z-a_j)} \hbox{e}^{-\kappa(z'-a_j)}+\frac{t_j^H\ \hbox{e}^{-\kappa a} \ r_i^H \ \hbox{e}^{-\kappa a} \ t_j^H}{2 \kappa \Delta_{12}^H} \hbox{e}^{-\kappa(z-a_j)} \hbox{e}^{-\kappa(z'-a_j)},
  && a_i<a_j<z,z^\prime.
  \end{align}
  \end{subequations}
  Region $g^{2,H}_{ \rput(0.1,0){\scriptstyle{11}}
  \pscircle[linewidth=0.2pt](0.1,0.01){0.15} }$ gives the transmission amplitude of the propagator across the interfaces with $\lambda^{\perp}_{e i}, \lambda^{\perp}_{g i}$ at $z=a_i$ and $\lambda^{\perp}_{e j}, \lambda^{\perp}_{g j}$ at $z=a_j$. The two terms in region $g^{2,H}_{ \rput(0.1,0){\scriptstyle{12}}
  \pscircle[linewidth=0.2pt](0.1,0.01){0.15} }$; the first term refers to the transmission of the propagator across the interface at $z=a_j$, and the second term represents the propagator reflected from the interface at $z=a_i$, with an exponential dependence on the length of propagation $a=(a_j-a_i)$ and then transmission across the interface at $z=a_j$. The three terms in region $g^{2,H}_{ \rput(0.1,0){\scriptstyle{13}}
  \pscircle[linewidth=0.2pt](0.1,0.01){0.15} }$; the first term refers to the propagator directly from the source, the second term refers to the reflection of the propagator across the interface at $z=a_j$ to the observation point, and the third term represents the propagator transmitted at the interface $z=a_j$, exponential dependence on the length of propagation $a$, reflected at interface $z=a_i$, again exponential dependence on the length of propagation $a$ and then transmission across the interface at $z=a_j$.
  
Casimir energies of $N=2,3,4,5$ plates were computed from the multiple scattering formalism using the reflection coefficients from $B^H$ of $g^{H}_{\rput(0.1,0){\scriptstyle{ij}}
 \pscircle[linewidth=0.2pt](0.1,0.01){0.15}}$ \cite{Abhignan_2023}. The reflection coefficients of $N=1$ were used in the Casimir energy calculation for $N=2$ setup. The reflection coefficients of $N=1$ and $N=2$ configurations were used in the Casimir energy calculation for $N=3$ plates. The Casimir energy for $N=4$ was calculated using the reflection coefficients of $N=1$ and $N=3$ plate configurations. Similarly, reflection coefficients of $N=2$ and $N=3$ configurations were utilized to get the Casimir energy for $N=5$ plates. Further, the corresponding Casimir force on $z=a_j$ plate in $N=2$ configuration was calculated from stress tensor method utilising regions $B^H_{13}$, $B^H_{22}$ in $g^{2,H}_{ \rput(0.1,0){\scriptstyle{ij}}\pscircle[linewidth=0.2pt](0.1,0.01){0.15} }$. The Casimir force on $z=a_k$ plate in $N=3$ configuration was calculated from stress tensor method utilising regions $B^H_{14}$, $B^H_{23}$ in $g^{3,H}_{ \rput(0.1,0){\scriptstyle{ij}}\pscircle[linewidth=0.2pt](0.1,0.01){0.15} }$.
 
We now see that the reciprocity theorem is satisfied by the regions of Green's functions where $z=z'$ ($i+j-2=N$ in Eq. \ref{matrix}), which were primarily utilized for Casimir calculations from multiple scattering formalism and stress tensor method \cite{Abhignan_2023}. Relying on previous studies \cite{Shajesh2011}, we observe the generalization for these regions 
\begin{equation}
 g^{N,H}_{i+j-2=N}(z,z^\prime) = g_0(z-z^\prime) 
+\tilde {\bf R}(z)^T \cdot {\bf t}^H_{i\ldots N} \cdot \tilde {\bf R}(z^\prime),
\label{12-rtr}
\end{equation}
with the relation between the free Green's function when all the plates are absent \begin{equation}
g_0(z-z^\prime) = \frac{1}{2\kappa} e^{-\kappa |z-z^\prime|}
\end{equation} and transition matrix ${\bf t}^H_{1\ldots N}$ of $N$ plates which are purely dielectric $\lambda^{\perp}_{g }\rightarrow0$ or purely permeable $\lambda^{\perp}_{e}\rightarrow0$ (Eq. \ref{coeff}). 

The vector ${\bf R}(z)$ is constructed out of the dimensionless free Green's function as
\begin{equation}
\tilde {\bf R}(z)^T = \left[\begin{array}{c c c}
 \tilde R_i(z) & \tilde R_j(z) & \cdots \end{array}\right] 
= \left[\begin{array}{c c c}
  e^{-\kappa |z-a_i|} & e^{-\kappa |z-a_j|} & \cdots \end{array}\right].
\label{vecrz}
\end{equation}
Further, utilizing the definitions
\begin{equation}
\quad
\tilde {\bf t}^H_{1\ldots N} = 2\kappa{\bf t}^H_{1\ldots N}
\quad \text{and} \quad \tilde {\bf R'} = 2\kappa {\bf R'}
\label{dimless-def}
\end{equation}
where \begin{equation}
{\bf R'} = \left[ \begin{array}{cccc}
g_0(0) & g_0(a_i-a_j) & \ldots & g_0(a_i-a_N) \\
g_0(a_j-a_i) &g_0(0) & \ldots & g_0(a_j-a_N) \\
\vdots & \vdots & \ddots & \vdots \\
g_0(a_N-a_i) &g_0(a_N-a_j) & \ldots & g_0(0)
\end{array} \right]
\label{Rm-def}
\end{equation}
we obtain \begin{equation}
\tilde {\bf t}^H_{i} = r_{i}
\end{equation}
for $N=1$ corresponding to $g^{1,H}_{i+j=3}(z,z^\prime)$ in Eq. \ref{matrix} ($z-z'$ space with regions $i+j=3$ in Fig. (\ref{1a})),
\begin{equation}
\tilde {\bf t}^H_{ij} = \frac{1}{\Delta_{ij}}
\left[ \begin{array}{cc}
r_i & r_i \tilde R'_{ij} \tilde r_j \\
\tilde r_j \tilde R'_{ji} \tilde r_i & \tilde r_j 
\end{array} \right] 
\end{equation} 
for $N=2$ corresponding to $g^{2,H}_{i+j=4}(z,z^\prime)$ in Eq. \ref{matrix} ($z-z'$ space with regions $i+j=4$ in Fig. (\ref{1b})) and
\begin{equation}
\tilde {\bf t}^H_{ijk} = \frac{1}{\Delta_{ijk}}
\left[ \begin{array}{ccc}
r_i(1-r_jr_k\tilde R'_{jk})
&r_i\tilde R'_{ij[k]}r_j &r_i\tilde R'_{ik[j]}r_k \\[2mm]
r_j\tilde R'_{ji[k]}r_i & r_j(1-r_kr_i\tilde R'_{ki})
&r_j\tilde R'_{jk[i]}r_k \\[2mm]
r_k\tilde R'_{ki[j]}r_i &r_k\tilde R'_{kj[i]}r_j 
&r_k(1-r_ir_j\tilde R'_{ij})
\end{array}\right]
\label{t123=expr}
\end{equation} for $N=3$ corresponding to $g^{3,H}_{i+j=5}(z,z^\prime)$ in Eq. \ref{matrix} ($z-z'$ space with regions $i+j=5$ in Fig. (\ref{Fig2}))
with
\begin{equation}
\tilde R'_{ij[k]} =\tilde R'_{ij}+\tilde R'_{ik} r_k \tilde R'_{kj},
\end{equation} for purely permeable plates $\lambda^{\perp}_{e i}\rightarrow0$ with $t^H_i=1+r^H_i$ in Eq. \ref{coeff} for $i,j,k$ plates. Similarly, we can obtain a transition matrix ${\bf t}^E_{1\ldots N}$ for TE mode Green's function $g^{N,E}_{i+j-2=N}(z,z^\prime)$ for purely dielectric plates $\lambda^{\perp}_{g i}\rightarrow0$ with $t^E_i=1+r^E_i$ in Eq. \ref{coeff} for $i,j,k$ plates. This representation in Eq. (12) is the Faddeev-like equation where the transition matrix ${\bf t}^H_{1\ldots N}$ decouples from the ${\bf R}$-vector and ${\bf t}^H_{1\ldots N}$ of $N$ plates can be solved recursively in terms of ${\bf t}^H_{1\ldots N-2}$ of $N-2$ plates \cite{selfsimilar}. 

\subsection{Casimir energy of $N$ magnetodielectric $\delta$-plates} 
It was observed that the multiple scattering parameter $\Delta_{ij\cdots N}$ can represent the Casimir energy $\Delta E_{(ij\cdots N)}$ of $N$ plates configuration as \cite{Abhignan_2023}, \begin{equation}
 \frac{\Delta E_{(ij\cdots N)}}{A} =  \frac{1}{2} \int_{-\infty}^\infty \frac{d\zeta}{2\pi}
\int \frac{d^2k_\perp}{(2\pi)^2} \Bigg[ \ln \Big[ \Delta_{ij\cdots N}^H \Big] + \ln \Big[ \Delta_{ij\cdots N}^E \Big] \Bigg].
\end{equation} The parameter $\Delta_{ij\cdots N}$ can be distributed into nearest neighbour scattering parameter $\Delta_{ij}$ for all $j=i+1$ ($i\in[1,N-1]$ where $i$ and $j$ are adjacent plates) \begin{equation}
    \Delta_{ij}= 1-r_i \hbox{e}^{-\kappa l_{ij}} r_j \hbox{e}^{-\kappa l_{ij}}, \label{nn}
\end{equation} and next-to-nearest neighbour, next-to-next-to-nearest neighbour, $\cdots$ scattering parameter  $\Delta_{ik}$ for all $k\geq i+2$ ($i\in[1,N-2]$ where $i$ and $k$ are not adjacent plates) \begin{equation}
    \Delta_{ik}= -r_i \hbox{e}^{-\kappa l_{i,i+1}} t_{i+1} \hbox{e}^{-\kappa l_{i+1,i+2}} t_{i+2} \cdots \hbox{e}^{-\kappa l_{k-1,k}} r_k \hbox{e}^{-\kappa l_{k-1,k}} \cdots t_{i+1} \hbox{e}^{-\kappa l_{i,i+1}}.
\end{equation} 
The scattering parameters show the various ways the propagation can contribute to the energy between the multiple plates. The Casimir energy in Eq. (21) for multiple plates is obtained from multiple scattering parameters \begin{subequations}
    \begin{align}
    N=2:&\Delta_{ij},\\N=3:&\Delta_{ijk}=\Delta_{ij}\Delta_{jk}+\Delta_{ik},\\N=4:&\Delta_{ijkl}=\Delta_{ij}\Delta_{jk}\Delta_{kl}+\Delta_{ij}\Delta_{jl}+\Delta_{ik}\Delta_{kl}+\Delta_{il},\\N=5:&\Delta_{ijklm}=\Delta_{ij}\Delta_{jk}\Delta_{kl}\Delta_{lm}+\Delta_{ij}\Delta_{jl}\Delta_{lm}+\Delta_{ik}\Delta_{kl}\Delta_{lm}+\Delta_{ij}\Delta_{jk}\Delta_{km}+\Delta_{ik}\Delta_{km}+\Delta_{ij}\Delta_{jm}+\Delta_{il}\Delta_{lm}+\Delta_{im},\\ &\cdots.
    \end{align}
    \end{subequations}
    The distribution of these multiple scattering parameters to nearest neighbour scattering parameter $\Delta_{ij}$ in Eq.(22) and next-to-nearest neighbour, next-to-next-to-nearest neighbour, $\cdots$ scattering parameter  $\Delta_{ik}$ in Eq.(23) of $N$ plates can be related to partitions and combinations of $N-1$ such as \begin{subequations}
    \begin{align}
    N-1=1:&\{1\}\implies\Delta_{ij},\\N-1=2:&\{1,1\}\implies\Delta_{ij}\Delta_{jk},\{2\}\implies\Delta_{ik},\\N-1=3:&\{1,1,1\}\implies\Delta_{ij}\Delta_{jk}\Delta_{kl},\{1,2\}\implies\Delta_{ij}\Delta_{jl},\{2,1\}\implies\Delta_{ik}\Delta_{kl},\{3\}\implies\Delta_{il},\\N-1=4:&\Biggl\{\begin{aligned}\{1,1,1,1\}\implies\Delta_{ij}\Delta_{jk}\Delta_{kl}\Delta_{lm},\{1,2,1\}\implies\Delta_{ij}\Delta_{jl}\Delta_{lm},\{2,1,1\}\implies\Delta_{ik}\Delta_{kl}\Delta_{lm},\\\{1,1,2\}\implies\Delta_{ij}\Delta_{jk}\Delta_{km},\{2,2\}\implies\Delta_{ik}\Delta_{km},\{1,3\}\implies\Delta_{ij}\Delta_{jm},\\\{3,1\}\implies\Delta_{il}\Delta_{lm},\{4\}\implies\Delta_{im},
    \end{aligned}\ \\ &\cdots.
    \end{align}
    \end{subequations}
    This behaviour shows that the number of combinations in partitions and their corresponding terms in the multiple scattering parameters for $N$ plates grows as $2^{(N-2)}$ with $N(N-1)/2$ independent terms such as \begin{align}
    N=2:&\Delta_{ij},\\N=3:&\Delta_{ij},\Delta_{jk},\Delta_{ik},\\N=4:&\Delta_{ij},\Delta_{jk},\Delta_{kl},\Delta_{ik},\Delta_{jl},\Delta_{il},\\N=5:&\Delta_{ij},\Delta_{jk},\Delta_{kl},\Delta_{lm},\Delta_{ik},\Delta_{jl},\Delta_{km},\Delta_{il},\Delta_{jm},\Delta_{im},\\ &\cdots.
    \end{align}
\section{Quasiperiodic sequence of magnetodielectric $\delta$-plates} 
The sequence of numbers $F_n = {1, 1, 2, 3, 5, 8, 13, 21,\cdots}$ called the Fibonacci series provides a basic concept of a quasi-periodic arrangement of numbers. Beginning with $F_1=F_2=1$, the consecutive terms in this series are derived from the recursive relation $F_{n+1} = F_n+F_{n-1}$. A mathematical relation between the Fibonacci series $F_n$ and the famous golden ratio $\phi$ is obtained by $\underset{n\gg1}{\text{lim}}F_{n+1}/F_{n}=\phi\approx1.618$ \cite{doi:10.1126/science.1081616}. 

A sequence of Fibonacci magnetodielectric $\delta$-plates is constructed with two different sorts of plates, $D'$ (perfectly conducting, $\lambda_{e}\rightarrow\infty$ in Eq. 9) and $N'$ (infinitely permeable,  $\lambda_{g}\rightarrow\infty$ in Eq. 9). We consider the substitution rules $D' \rightarrow D'N'$ and $N' \rightarrow D'$, when applied consecutively, produce a sequence of configurations \begin{equation}
   I=1: D'N' \rightarrow I=2: D'N'D' \rightarrow I=3: D'N'D'D'N' \rightarrow I=4: D'N'D'D'N'D'N'D' \rightarrow\cdots
\end{equation} with equal spacing $a$ between the plates and $I$ indicates the iteration. The Casimir energy of these ideal configurations of plates in Eq. (21) becomes \begin{equation}
\frac{\Delta E_{(ij\cdots N)}}{A} = \frac{1}{2} \int_{-\infty}^\infty \frac{d\zeta}{2\pi}
\int \frac{d^2k_\perp}{(2\pi)^2} \Bigg[ \ln \Big[ \Delta_{ij}^H \Delta_{jk}^H \cdots \Delta_{N-1,N}^H\Big] + \ln \Big[ \Delta_{ij}^E \Delta_{jk}^E \cdots \Delta_{N-1,N}^E\Big] \Bigg],
\end{equation} where the multiple $\Delta_{ijk\cdots N}$ scattering parameters become 
\begin{equation}
    \Delta_{ijk\cdots}= (1-r_i \hbox{e}^{-\kappa a} r_j \hbox{e}^{-\kappa a})(1-r_j \hbox{e}^{-\kappa a} r_k \hbox{e}^{-\kappa a})\cdots,
\end{equation} where optical coefficients of $\delta$ plate with an infinitely permeable material ($\lambda_{gi} \rightarrow \infty$, $\lambda_{ei}=0$) having $r^H=-1,r^E=1$ and infinitely dielectric material ($\lambda_{ei} \rightarrow \infty$, $\lambda_{gi}=0$) having $r^H=1,r^E=-1$ from Eq. 9. The parameter $\Delta_{ijk\cdots}$ in this case only consists of the nearest neighbour scattering parameters $\Delta_{ij}$, $\Delta_{jk}$ and so on for ideal magnetodielectric plates without the next-to-nearest neighbour, next-to-next-to-nearest neighbour, $\cdots$ scattering parameters. Using this, we obtain the Casimir energy for the sequence of plates in Eq. (31) as
\begin{subequations}
\begin{align}
   I=1:&\Delta E_{(D'N')} = -\frac{7}{8}\Delta E_{(D'D')},\\
   I=2:& \Delta E_{(D'N'D')} = -\frac{7}{8}\Delta E_{(D'D')}-\frac{7}{8}\Delta E_{(D'D')},\\
    I=3:&\Delta E_{(D'N'D'D'N')} = -\frac{7}{8}\Delta E_{(D'D')}-\frac{7}{8}\Delta E_{(D'D')}+\Delta E_{(D'D')}-\frac{7}{8}\Delta E_{(D'D')},\\
   I=4:&\Biggl\{\begin{aligned} \Delta E_{(D'N'D'D'N'D'N'D')} = -\frac{7}{8}\Delta E_{(D'D')}-\frac{7}{8}\Delta E_{(D'D')}+\Delta E_{(D'D')}-\frac{7}{8}\Delta E_{(D'D')}-\frac{7}{8}\Delta E_{(D'D')}\\-\frac{7}{8}\Delta E_{(D'D')}-\frac{7}{8}\Delta E_{(D'D')},
  \end{aligned}\\
   \cdots
\end{align}
\end{subequations}
where Casimir energy between two perfectly conducting plates is $\Delta E_{(D'D')}/A = -\pi^2/720a^3$ which leads to an attractive force\,\cite{Casimir:1948dh} (spectral distribution of Bose-Einstein statistics \cite{Brevik2018}), whereas in Eq. (34a) $\Delta E_{(D'N')}/A = 7/8(\pi^2/720a^3)$ is Casimir energy of Boyer configuration which leads to repulsive force\,\cite{Boyer1974} (spectral distribution of Fermi-Dirac statistics \cite{Brevik2018}). 

After $I=25$ iterations we numerically find all these structures have positive sign for Casimir interaction energy which grow geometrically as $\Delta E_{(D'N'\cdots)}\sim0.51\text{e}^{0.48 I}|\Delta E_{(D'D')}|$ and as a result, the vacuum pressure causes the sequence of plates in Fibonacci configurations to inflate. The Casimir energy of consecutive sequence of configurations approaches $\Delta E_{(D'N'\cdots)}(I+1)/\Delta E_{(D'N'\cdots)}(I)\approx1.618$, golden ratio $\phi$. Similarly, the number of plates $N$ increases geometrically in this Fibonacci sequence as $N\sim1.2\text{e}^{0.48 I}$. We also numerically study the number of neighbours at which the corresponding Casimir energies differ to identify whether the stack of plates in a quasiperiodic sequence expands or contracts. The number of neighbourhoods with $D'D'$ or $N'N'$, $D'N'$ or $N'D'$ grows for the configuration of plates in the Fibonacci sequence as $N(D'D')+N(N'N')\sim0.28\text{e}^{0.48 I}$, $N(N'D')+N(D'N')\sim0.89\text{e}^{0.48 I}$, respectively. These reveal that the stack will expand for a finite number of iterations since it has more repulsive neighbourhoods than attractive neighbourhoods.

  \begin{figure}[!hb]
\centering
\begin{subfigure}{0.8\linewidth}
\includegraphics[width=\linewidth]{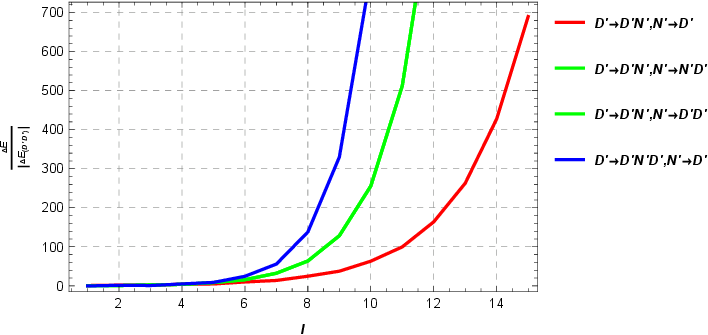}
\caption{Casimir energy scaled by $|\Delta E_{(D'D')}|$ for quasiperiodic sequences which expand at different iterations $I$.}
\end{subfigure}
\begin{subfigure}{0.8\linewidth}
\includegraphics[width=1\linewidth]{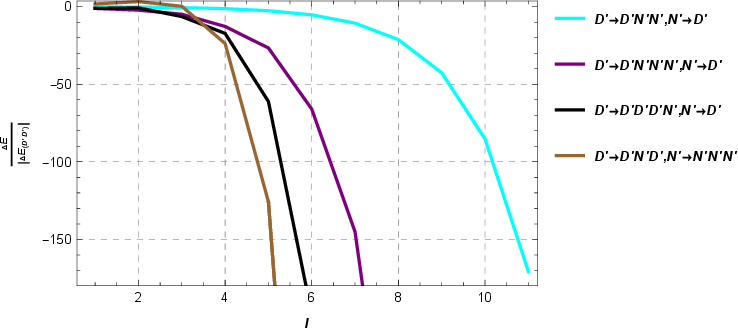}
\caption{Casimir energy scaled by $|\Delta E_{(D'D')}|$ for quasiperiodic sequences which contract at different iterations $I$.}
\end{subfigure}
\caption{Casimir energy of quasiperiodic sequences with perfectly conducting and permeable $\delta$-plates obtained from substitution rules \cite{Maciá_2006}.}
\end{figure}

\begin{table}
\caption{Substitution rules that determine the sequences of quasiperiodic structures \cite{Maciá_2006}, their Casimir energies scaled by $|\Delta E_{(D'D')}|$ after $I$ iterations and the number of plates $N$ in the sequence.}
\label{tab:substitution_rules}
\begin{tabular}{|c|c|c|c|c|c|c|}
\hline
\textbf{Sequence} & \textbf{Substitution rules} &\begin{tabular}{c c} \textbf{Casimir energies} \\  ($\Delta E$/$|\Delta E_{(D'D')}|$$\sim$)  \end{tabular} & \scriptsize $N\sim$ & \scriptsize $N(D'D',N'N')\sim$& \scriptsize $N(D'N',N'D'$)$\sim$& $\frac{\Delta E(I+1)}{\Delta E(I)}\approx$\\
\hline
Fibonacci  & $D'\rightarrow D'N', N' \rightarrow D'$ & $0.51\text{e}^{0.48 I}$ & $1.2\text{e}^{0.48 I}$ & $0.28\text{e}^{0.48 I}$ & $0.89\text{e}^{0.48 I}$ & 1.618\\
\hline
Thue-Morse  & $D'\rightarrow D'N', N' \rightarrow N'D'$ & $0.25\text{e}^{0.69 I}$ & $1.0\text{e}^{0.69 I}$ & $0.33\text{e}^{0.69 I}$ & $0.67\text{e}^{0.69 I}$ & 2\\
\hline
Period-doubling  & $D'\rightarrow D'N', N' \rightarrow D'D'$ & $0.25\text{e}^{0.69 I}$ & $1.0\text{e}^{0.69 I}$ & $0.33\text{e}^{0.69 I}$ & $0.67\text{e}^{0.69 I}$ & 2\\
\hline
Silver mean  & $D'\rightarrow D'N'D', N' \rightarrow D'$ & $0.12\text{e}^{0.88 I}$ & $1.2\text{e}^{0.88 I}$ & $0.50\text{e}^{0.88 I}$ & $0.71\text{e}^{0.88 I}$ & 2.414\\
\hline
Bronze mean  & $D'\rightarrow D'D'D'N', N' \rightarrow D'$ & $-0.15\text{e}^{1.2 I}$ & $1.2\text{e}^{1.2 I}$ & $0.64\text{e}^{1.2 I}$ & $0.56\text{e}^{1.2 I}$ & 3.302\\
\hline
Copper mean  & $D'\rightarrow D'N'N', N' \rightarrow D'$ & $-0.08\text{e}^{0.69 I}$ & $1.3\text{e}^{0.69 I}$ & $0.67\text{e}^{0.69 I}$ & $0.67\text{e}^{0.69 I}$ & 2\\
\hline
Nickel mean  & $D'\rightarrow D'N'N'N', N' \rightarrow D'$ & $-0.43\text{e}^{0.83 I}$ & $1.47\text{e}^{0.83 I}$ & $0.92\text{e}^{0.83 I}$ & $0.56\text{e}^{0.83 I}$ & 2.302\\
\hline
Triadic Cantor  & $D'\rightarrow D'N'D', N' \rightarrow N'N'N'$ & $-0.79\text{e}^{1.1 I}$ & $1.0\text{e}^{1.1 I}$ & $0.88\text{e}^{1.1 I}$ & $1.9\text{e}^{0.69 I}$ & 3\\
\hline
\end{tabular}
\end{table}

Similarly, we compute Casimir energies for a class of quasiperiodic sequences, usually considered in the study of self-similar systems where the sequences are defined in terms of specific substitution rules \cite{Maciá_2006}, as shown in Table I. Depending on the substitution rule, the stack of perfectly conducting and infinitely permeable plates in the quasiperiodic sequence expands or contracts with positive or negative Casimir energies. It is also evident that growth in the number of repulsive neighbourhoods $D'N', N'D'$ (Fermi-Dirac distribution) and attractive neighbourhoods $D'D',N'N'$ (Bose-Einstein distribution) indicates if the stack expands or contracts. Further, we illustrate the growth of Casimir energies depending on their substitution rule for the sequences that expand in Fig. 3. (a) and sequences that contract in Fig. 3. (b). This behavior, as numerically computed, is only observed for finite sequences with finite iterations; for an infinite sequence, it may change \cite{selfsimilar} where it was observed in an exact manner that self-similar structures described by scalar field with neighboring repulsive behavior could contract at the infinite limit.

\begin{figure}[!hb]
\centering
\begin{subfigure}{0.75\linewidth}
\includegraphics[width=\linewidth]{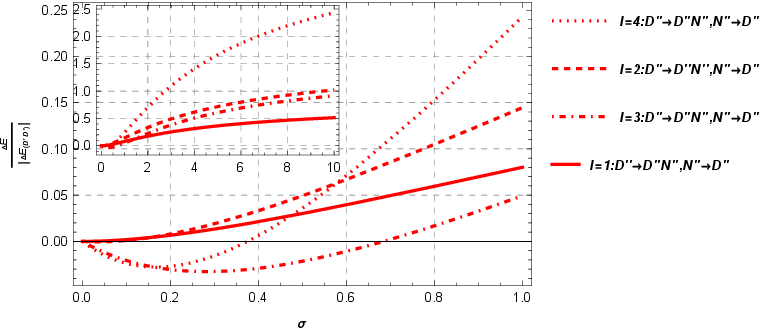}
\caption{Casimir energy scaled by $|\Delta E_{(D'D')}|$ by varying dispersion $\sigma$ for Fibonacci sequence at iterations $I=1,2,3,4$ obtained from substitution rule $D''\rightarrow D''N'', N'' \rightarrow D''$.}
\end{subfigure}
\begin{subfigure}{0.75\linewidth}
\includegraphics[width=1\linewidth]{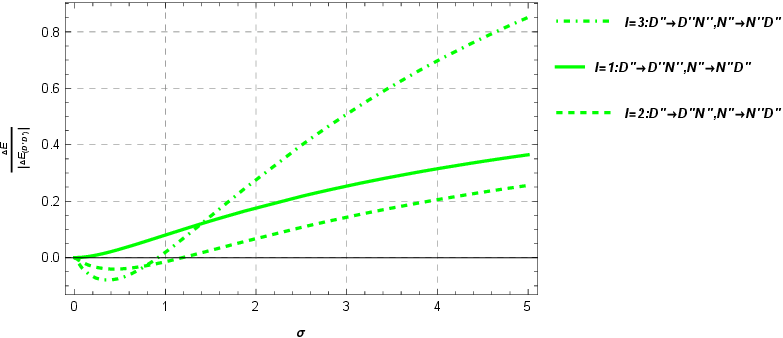}
\caption{Casimir energy scaled by $|\Delta E_{(D'D')}|$ by varying dispersion $\sigma$ for Thue-Morse sequence at iterations $I=1,2,3$ obtained from substitution rule $D''\rightarrow D''N'', N'' \rightarrow N''D''$.}
\end{subfigure}
\caption{Casimir energy of quasiperiodic sequences which expand with finitely conducting and permeable $\delta$-plates}
\end{figure}

Further, we numerically study the Casimir energies with finite material properties to observe if these quasiperiodic sequences expand in a similar manner. To study Casimir energies of finite material properties, we consider constant dispersion for electric and magnetic properties of plates such as  \begin{equation}
     \lambda_{e}^{\perp}(\zeta)= \lambda_{g}^{\perp}(\zeta)=\frac{\sigma}{\zeta}. \end{equation} We consider the same constant and isotropic optical conductivity $\sigma$ for dielectric and permeable materials in Eq. (9) so that the sequence is quasiperiodic with the same properties at all frequencies. With this, the optical coefficients are \begin{align}
        r_{D''}^H= \frac{\sigma \kappa}{\sigma \kappa +2 \zeta}, \ r_{D''}^E= -\frac{\sigma \zeta}{\sigma \zeta+2 \kappa}, \ t^H_{D''}=1-r^H_{D''}, \ t^E_{D''}=1+r^E_{D''}
         \end{align}  for purely dielectric plate  $D''(\lambda_{ei}^{\perp}= \sigma/\zeta,\lambda^{\perp}_{g i}\rightarrow0)$ and
         \begin{equation}
          r_{N''}^H= -\frac{\sigma \zeta}{\sigma \zeta +2 \kappa}, \ r_{N''}^E=\frac{\sigma \kappa}{\sigma \kappa+2 \zeta}, \ t^H_{N''}=1+r^H_{N''}, \ t^E_{N''}=1-r^E_{N''}
          \end{equation} for purely permeable plate $N'' (\lambda_{gi}^{\perp}= \sigma/\zeta,\lambda^{\perp}_{e i}\rightarrow0)$ from Eq. (9). Using these, we compute the Casimir energy for the Fibonacci sequence ($D''\rightarrow D''N'', N'' \rightarrow D''$) of plates with finite material properties as
          \begin{subequations}
\begin{align}
   I=1:&\Delta E_{(D''N'')} =  \frac{1}{2} \int_{-\infty}^\infty \frac{d\zeta}{2\pi}
\int \frac{d^2k_\perp}{(2\pi)^2} \Bigg[ \ln \Big[1 - r^H_{D''}r^H_{N''} e^{-2 \kappa a}\Big] + \ln \Big[1 - r^E_{D''}r^E_{N''} e^{-2 \kappa a}\Big] \Bigg],\\
   I=2:&\Biggl\{\begin{aligned}
       \Delta E_{(D''N''D'')} = \frac{1}{2} \int_{-\infty}^\infty \frac{d\zeta}{2\pi}
\int \frac{d^2k_\perp}{(2\pi)^2} \Bigg[ \ln \Big[(1 - r^H_{D''}r^H_{N''} e^{-2 \kappa a})(1 - r^H_{N''}r^H_{D''} e^{-2 \kappa a})-(r^H_{D''})^2(t^H_{N''})^2 e^{-4 \kappa a})\Big] \\+ \ln \Big[(1 - r^E_{D''}r^E_{N''} e^{-2 \kappa a})(1 - r^E_{N''}r^E_{D''} e^{-2 \kappa a})-(r^E_{D''})^2(t^E_{N''})^2 e^{-4 \kappa a})\Big]\Bigg],\end{aligned}\\
   \cdots
\end{align}
\end{subequations}
using Eq. (24) in Eq. (21). We describe this in detail in Appendix B.

 The Casimir energy is numerically evaluated and scaled by $|\Delta E_{(D'D')}|$ for varying dispersion $\sigma$ values at different iterations $I=1,2,3,4$ as displayed in Fig. 4. (a). We observe that the Fibonacci sequence of plates contract for small values of $\sigma$, and as $\sigma$ increases, the plates expand. As the value of $\sigma$ increases further increases, it can be seen in the inset of Fig. 4. (a) that the values for Casimir energy saturate to ideal conditions  ($\sigma \rightarrow \infty$) as in Eq. (32). Similar behaviour is observed for Casimir energy for Thue-Morse sequence of plates with purely dielectric and permeable plates ($D''\rightarrow D''N'', N'' \rightarrow N''D''$) by varying $\sigma$ values at different iterations $I=1,2,3$ as displayed in Fig. 4. (b).

\section{Conclusion}

We examined finite configurations of quasiperiodic structures made from magnetodielectric $\delta$-function plates that are equally spaced. The Casimir energy for a class of quasiperiodic structures constructed from purely conducting or permeable $\delta$-plates is positive or negative, making the stack expand or contract depending on the neighbourhoods. We numerically find the relation between Casimir energies and the number of plates in the quasiperiodic structures for finite-size lattices. The Casimir energy expression for $N$ $\delta$-plates plates was utilized \cite{Abhignan_2023}, and a general prescription for numerically evaluating the energies with constant and isotropic conductivity for the dielectric or magnetic properties of $N$ $\delta$-plates is described here. 

We also handled Green's functions for the magnetodielectric $\delta$-plates and used the scattering matrix approach \cite{Shajesh2011} to construct the Faddeev-like equation for $N$ purely conducting or permeable plates. While previously only $N=1,2,3$ $\delta$-plates were handled \cite{Abhignan_2023}, here we could generalize the expression of Green's functions for $N$ plates with dielectric or magnetic properties in the regions where $z=z'$. Faddeev's equations were primarily known for nuclear many-body scattering \cite{faddeev1993quantum,Faddeev1965} and utilized previously in optical studies \cite{PhysRev.92.291,PhysRev.97.1344}. The Martin-Schwinger-Puff many-body theory is also closely related \cite{PhysRev.115.1342,PUFF1961}. Expanding the $N$-body scattering matrix for Green's functions in all the regions with both dielectric and magnetic properties may be more interesting. 

 In Casimir's seminal work \cite{Casimir:1948dh}, the attractive force was derived between two plates with dielectric permittivity $\epsilon_1 \rightarrow \infty$ and $\epsilon_2 \rightarrow \infty$. Lifshitz et al. generalized the Casimir result for two dielectric slabs with arbitrary values of $\epsilon_1$ and $\epsilon_2$ with $\epsilon_3$ in the space filled in-between them \cite{LifshitzAIP}. This result gives rise to only attractive force when we consider similar bodies $\epsilon_1=\epsilon_2\equiv\epsilon$ regardless of $\epsilon_3$. However, it can turn repulsive when a dissimilarity arises, such as  $\epsilon_1>\epsilon_3>\epsilon_2$ in the dielectric response functions, which was experimentally verified recently \cite{Munday2009}. Beyond the Casimir-Lifshitz result only using dielectrics, Casimir repulsion can also be recognized in several other contexts of standard electrodynamics, such as using magnetically permeable materials by Kenneth et al. \cite{RCF}, similar to Boyer \cite{Boyer1974} and from changing geometries by Levin et al. \cite{Levin2010}. Repulsion is essential in nano- and micro-structures where the Casimir forces predominate, leading to stiction \cite{PhysRevB.63.033402,PhysRevLett.87.211801}. Boyer repulsion is an intriguing solution \cite{RCF}; yet, it has been considered non-physical and difficult to manifest as naturally existing materials do not show significant magnetic responses\,\cite{CRCF,PhysRevLett.91.029102}. Nevertheless, metamaterials with magnetic responses have been created recently through breakthroughs in nanofabrication\,\cite{Shalaev2007}, which could be pertinent to the physical realization of Boyer's repulsion \cite{PhysRevA.80.042510,PhysRevLett.103.120401}. And the weak Boyer repulsion could be magnified with a quasiperiodic arrangement.

 \pagebreak
 
\appendix
\section{Green's functions of $N$ $\delta$-plates}
 The matrices $A$, $B^H$, and $C$ representing distinct regions in $z-z'$ space for Green's function $g^{2,H}$ of $N=2$ with the subscript label in Fig. (\ref{1b}) are
 \begin{multline}
     A = \left[\begin{array}{c c c}
  \hbox{e}^{-\kappa(z'-a_j)} &  \left[\begin{array}{c c}
   \hbox{e}^{-\kappa(z'-a_i)} &  \hbox{e}^{-\kappa(a_j-z')} \end{array}\right] &  \hbox{e}^{-\kappa(a_i-z')} \end{array}\right], \\ B^H = \left[\begin{array}{c c c} \frac{t_i^H \hbox{e}^{-\kappa a} t_j^H}{\Delta_{ij}^H} & \left[\begin{array}{c c}
   \frac{t_j^H}{\Delta_{ij}^H} \hspace{1.5mm} & \hspace{1.5mm} \frac{r_i^H \hbox{e}^{-\kappa a} t_j^H}{\Delta_{ij}^H} \end{array}\right] &  r_j^H +\frac{t_j^H\hbox{e}^{-\kappa a}r_i^H\hbox{e}^{-\kappa a}t_j^H}{\Delta_{ij}^H} \vspace{3mm}  \\ \left[\begin{array}{c c}
  \frac{t_i^H}{\Delta_{ij}^H} \vspace{3mm} \\ \vspace{3mm}\frac{t_i^H \hbox{e}^{-\kappa a} r_j^H}{\Delta_{ij}^H} \end{array}\right] & \left[\begin{array}{c c} \frac{r_i^H \hbox{e}^{-\kappa a} r_j^H}{\Delta_{ij}^H} & \frac{r_i^H}{\Delta_{ij}^H} \vspace{3mm} \\ \vspace{3mm} \frac{r_j^H}{\Delta_{ij}^H} & \frac{r_j^H \hbox{e}^{-\kappa a} r_i^H}{\Delta_{ij}^H}
  \end{array}\right] & \left[\begin{array}{c c}
   \frac{r_i^H \hbox{e}^{-\kappa a} t_j^H}{\Delta_{ij}^H} \vspace{3mm} \\ \vspace{3mm} \frac{t_j^H}{\Delta_{ij}^H} \end{array}\right] \vspace{3mm} \\ \vspace{3mm} r_i^H +\frac{t_i^H\hbox{e}^{-\kappa a}r_j^H\hbox{e}^{-\kappa a}t_i^H}{\Delta_{ij}^H} &\left[\begin{array}{c c}
  \frac{r_j^H\hbox{e}^{-\kappa a}t_i^H}{\Delta_{ij}^H} &  \frac{t_i^H}{\Delta_{ij}^H} \end{array}\right] &  \frac{t_j^H \hbox{e}^{-\kappa a} t_i^H}{\Delta_{ij}^H}
  \end{array}\right]\\\hbox{and}\, C = \left[\begin{array}{c c c}
  \hbox{e}^{-\kappa(a_i-z)} \\ \left[\begin{array}{c c}
  \hbox{e}^{-\kappa(a_j-z)} \\ \hbox{e}^{-\kappa(z-a_i)} \end{array}\right] \\  \hbox{e}^{-\kappa(z-a_j)} \end{array}\right].
  \end{multline}
  Green's function $ g^{2,H}_{ \rput(0.1,0){\scriptstyle{ij}}
 \pscircle[linewidth=0.2pt](0.1,0.01){0.15} }$ in various regions of $z-z'$ space can be obtained here from Eq. (\ref{matrix}) for $N=2$. Here, \begin{equation}
    \Delta_{ij}^H=1-r_i^H \hbox{e}^{-\kappa a} r_j^H \hbox{e}^{-\kappa a},
    \label{MS1}
 \end{equation} is the multiple scattering parameter. The distance between the plates $i$ and $j$ is $a=a_j-a_i$.
 
 The matrices $A$, $B^H$, and $C$ representing distinct regions in $z-z'$ space for Green's function $g^{3,H}$ of $N=3$ with the subscript label in Fig. (\ref{Fig2}) are
\begin{multline}
     A = \left[\begin{array}{c c c c}
\hbox{e}^{-\kappa(z'-a_k)} &  \left[\begin{array}{c c} \hbox{e}^{-\kappa(z'-a_j)} &  \hbox{e}^{-\kappa(a_k-z')} \end{array}\right] &  \left[\begin{array}{c c}
  \hbox{e}^{-\kappa(z'-a_i)} &  \hbox{e}^{-\kappa(a_j-z')} \end{array}\right] &  \hbox{e}^{-\kappa(a_i-z')} \end{array}\right], \\ B^H = \frac{1}{\Delta_{ijk}^H} \left[\begin{array}{cccc}
 \substack{t_{i}^H \hbox{e}^{-\kappa a} t_{j}^H \hbox{e}^{-\kappa b} t_{k}^H }  &  B'_{12} & B'_{13} & \substack{ r_{k}^H \Delta_{ijk}^H + t_{k}^H \hbox{e}^{-\kappa b} r_{j}^H \hbox{e}^{-\kappa b} t_{k}^H \Delta_{ij}^H \\ + t_{k}^H \hbox{e}^{-\kappa b} t_{j}^H \hbox{e}^{-\kappa a} r_{i}^H \hbox{e}^{-\kappa a} t_{j}^H \hbox{e}^{-\kappa b} t_{k}^H}  \\ 
 B'_{21}   &  B'_{22} & B'_{23} & B'_{24} \\
 B'_{31}   &  B'_{32} & B'_{33} & B'_{34}  \\  \substack{
 r_i^H \Delta_{ijk}^H + t_i^H \hbox{e}^{-\kappa a} r_j^H \hbox{e}^{-\kappa a} t_i^H \Delta_{jk}^H  \\ + t_i^H \hbox{e}^{-\kappa a} t_j^H \hbox{e}^{-\kappa b} r_k^H \hbox{e}^{-\kappa b} t_j^H \hbox{e}^{-\kappa a} t_i^H}  &  B'_{42} & B'_{43} & \substack{ t_{k}^H \hbox{e}^{-\kappa b} t_{j}^H \hbox{e}^{-\kappa a} t_{i}^H }
 \end{array}\right] \\\hbox{and}\, C = \left[\begin{array}{c c c c}
 \hbox{e}^{-\kappa(a_i-z)} \\ \left[\begin{array}{c c}
 \hbox{e}^{-\kappa(a_j-z)} \\ \hbox{e}^{-\kappa(z-a_i)} \end{array}\right]  \vspace{3mm} \\  \vspace{3mm} \left[\begin{array}{c c}
 \hbox{e}^{-\kappa(a_k-z)}  \\  
  \hbox{e}^{-\kappa(z-a_j)} \end{array}\right] \\ \hbox{e}^{-\kappa(z-a_k)} \end{array}\right]
\end{multline}
 with
 \begin{equation}
     B'_{12}=\left[\begin{array}{c c} \substack{t_{j}^H  \hbox{e}^{-\kappa b} t_{k}^H} \hspace{3mm} & \hspace{3mm} \substack{ r_{i}^H \hbox{e}^{-\kappa a} t_{j}^H \hbox{e}^{-\kappa b} \ t_{k}^H } \end{array}\right],\,B'_{34}=\left[\begin{array}{c c} \substack{t_{k}^H \hbox{e}^{-\kappa b} t_{j}^H \hbox{e}^{-\kappa a} r_{i}^H} \\ \substack{t_{k}^H  \hbox{e}^{-\kappa b} t_{j}^H} \end{array}\right],
 \end{equation}
 \begin{equation}
    B'_{13}=\left[\begin{array}{c c}\substack{t_{k}^H \Delta_{ij}^H} \hspace{3mm} &  \hspace{3mm} \substack{r_{j}^H \hbox{e}^{-\kappa b}  t_{k}^H \Delta_{ij}^H + t_{j}^H\ \hbox{e}^{-\kappa a} r_{i}^H \hbox{e}^{-\kappa a} t_{j}^H \hbox{e}^{-\kappa b} t_{k}^H } \end{array}\right],\,B'_{24}=\left[\begin{array}{c c} \substack{ t_{k}^H \hbox{e}^{-\kappa b} r_{j}^H \Delta_{ij}^H + t_{k}^H \hbox{e}^{-\kappa b}  t_{j}^H  \hbox{e}^{-\kappa a}  r_{i}^H \hbox{e}^{-\kappa a} t_{j}^H } \\ \substack{t_{k}^H \Delta_{ij}^H } \end{array}\right],
 \end{equation}
 \begin{equation}
   B'_{21}=\left[\begin{array}{c c}
\substack{t_{i}^H \hbox{e}^{-\kappa a} t_{j}^H} \\ \substack{t_{i}^H \hbox{e}^{-\kappa a} t_{j}^H \hbox{e}^{-\kappa b} r_{k}^H} \end{array}\right],\,B'_{43}=\left[\begin{array}{c c} \substack{r_{k}^H \hbox{e}^{-\kappa b} t_{j}^H \hbox{e}^{-\kappa a} t_{i}^H} \hspace{3mm} & \hspace{3mm} \substack{ t_{j}^H \hbox{e}^{-\kappa a} t_{i}^H}\end{array}\right],
 \end{equation}
 \begin{equation}
     B'_{22}=\left[\begin{array}{c c} \substack{ t_{j}^H} & \substack{r_{i}^H \hbox{e}^{-\kappa a} t_{j}^H} \\ \substack{t_{j}^H \hbox{e}^{-\kappa b} r_{k}^H} & \substack{ r_{i}^H \hbox{e}^{-\kappa a} t_{j}^H \hbox{e}^{-\kappa b} r_{k}^H }
\end{array}\right],\,B'_{33}=\left[\begin{array}{c c} \substack{r_{k}^H \hbox{e}^{-\kappa b} t_{j}^H \hbox{e}^{-\kappa a} r_{i}^H} & \substack{t_{j}^H \hbox{e}^{-\kappa a} r_{i}^H} \\ \substack{r_{k}^H \hbox{e}^{-\kappa b} t_{j}^H} & \substack{t_{j}^H}
 \end{array}\right],
 \end{equation}
 \begin{equation}
     B'_{31}=\left[\begin{array}{c c}
 \substack{t_{i}^H \Delta_{jk}^H}\\ \substack{t_{i}^H \hbox{e}^{-\kappa a}  r_{j}^H  \Delta_{jk}^H + t_{i}^H \hbox{e}^{-\kappa a}  t_{j}^H  \hbox{e}^{-\kappa b}  r_{k}^H  \hbox{e}^{-\kappa b}  t_{j}^H} \end{array}\right],\,B'_{42}=\left[\begin{array}{c c} \substack{t_{i}^H \Delta_{jk}^H} \hspace{3mm} &  \hspace{3mm} \substack{r_{j}^H \hbox{e}^{-\kappa a}  t_{i}^H \Delta_{jk}^H +
 t_{j}^H \hbox{e}^{-\kappa b} r_{k}^H \hbox{e}^{-\kappa b} t_{j}^H \hbox{e}^{-\kappa a} t_{i}^H} \end{array}\right],
 \end{equation}
 \begin{equation}
     B'_{23}= \left[\begin{array}{c c} \substack{r_{k}^H \hbox{e}^{-\kappa b} r_{j}^H \Delta_{ij}^H + r_{k}^H \hbox{e}^{-\kappa b} t_{j}^H \hbox{e}^{-\kappa a} r_{i}^H \hbox{e}^{-\kappa a} t_{j}^H } \hspace{3mm} & \hspace{3mm} \substack{ r_{j}^H \Delta_{ij}^H+ t_{j}^H \hbox{e}^{-\kappa a} r_{i}^H \hbox{e}^{-\kappa a} t_{j}^H } \\ \substack{ r_{k}^H  \Delta_{ij}^H } & \substack{ t_{j}^H \hbox{e}^{-\kappa a} r_{i}^H \hbox{e}^{-\kappa a} t_{j}^H \hbox{e}^{-\kappa b} r_{k}^H+r_{j}^H \hbox{e}^{-\kappa b} r_{k}^H  \Delta_{ij}^H }
 \end{array}\right]
 \end{equation} and 
 \begin{equation}
     B'_{32}= \left[\begin{array}{c c} \substack{t_{j}^H \hbox{e}^{-\kappa b} r_{k}^H \hbox{e}^{-\kappa b} t_{j}^H \hbox{e}^{-\kappa a} r_{i}^H+r_{j}^H \hbox{e}^{-\kappa a} r_{i}^H \Delta_{jk}^H} & \substack{r_{i}^H  \Delta_{jk}^H} \\ \substack{r_{j}^H \Delta_{jk}^H + t_{j}^H \hbox{e}^{-\kappa b} r_{k}^H \hbox{e}^{-\kappa b} t_{j}^H} & \substack{ r_{i}^H \hbox{e}^{-\kappa a} r_{j}^H \Delta_{jk}^H + r_{i}^H \hbox{e}^{-\kappa a} t_{j}^H \hbox{e}^{-\kappa b} r_{k}^H \hbox{e}^{-\kappa b} t_{j}^H}
 \end{array}\right].
 \end{equation}
  Green's function $ g^{3,H}_{ \rput(0.1,0){\scriptstyle{ij}}
 \pscircle[linewidth=0.2pt](0.1,0.01){0.15} }$ in various regions of $z-z'$ space can be obtained here from Eq. (\ref{matrix}) for $N=3$. Here, \begin{equation}
              \Delta_{ijk}^H=(\Delta_{ij}^H\Delta_{jk}^H-r_i^H \hbox{e}^{-\kappa a}  t_j^H \hbox{e}^{-\kappa b} r_k^H \hbox{e}^{-\kappa b} t_j^H \hbox{e}^{-\kappa a})\,\hbox{and}\,\Delta_{jk}^H=1-r_j^H \hbox{e}^{-\kappa b} r_k^H \hbox{e}^{-\kappa b}, \label{MS3}
              \end{equation} 
                is the multiple scattering parameter. The distance between the plates $j$ and $k$ is $b=a_k-a_j$.

\section{Casimir energies of $N$ $\delta$-plates}
In general, the Casimir energy of $N=2$ $\delta$-plates with reflection coefficients $r_i,r_j=r_{D''(N'')}$ (Eqs. (36), (37)) from Eq. (21) is  \begin{equation}
 \frac{\Delta E_{(ij)}}{A} = \frac{1}{2} \int_{-\infty}^\infty \frac{d\zeta}{2\pi}
\int \frac{d^2k_\perp}{(2\pi)^2} \Bigg[ \ln \Big[1 - r^H_i r^H_j e^{-2 \kappa a}\Big] + \ln \Big[1 - r^E_i r^E_j e^{-2 \kappa a}\Big] \Bigg].
\end{equation} 
Considering the spherical polar coordinates $k_\perp = \kappa \ \hbox{sin} \theta$, $\zeta = \kappa \ \hbox{cos} \theta$\,\cite{Brevik2018} we obtain \begin{equation}
 \frac{\Delta E_{(ij)}}{|\Delta E_{(D'D')}|} =  \frac{45}{2\pi^4} 
\int_{0}^{1} dt \int_{0}^\infty s^2 ds \Bigg[ \ln \Big[1 - r^H_i r^H_j e^{- s}\Big] + \ln \Big[1 - r^E_i r^E_j e^{- s}\Big] \Bigg]
\end{equation} scaled by $\Delta E_{(D'D')}/A = -\pi^2/720a^3$ with the reflection coefficients in Eqs. (36), (37) as \begin{align}
        r_{D''}^H= \left(\frac{\sigma }{\sigma  +2 t}\right), \ r_{D''}^E= \left(-\frac{\sigma}{\sigma+\frac{2}{t}}\right), 
          r_{N''}^H= r_{D''}^E, \ r_{N''}^E= r_{D''}^H.
          \end{align}
Further, we can evaluate the $s$ integral to obtain \begin{equation}
 \frac{\Delta E_{(ij)}}{|\Delta E_{(D'D')}|} =  -\frac{45}{\pi^4} 
\int_{0}^{1} dt  \Bigg[ \text{Li}_4 \left(r^H_i r^H_j\right) + \text{Li}_4\left(r^E_i r^E_j\right) \Bigg]
\end{equation} where $\text{Li}_4(z)$ is a polylogarithm function and the $t$ integral can be numerically evaluated for varying $\sigma$ values.

Similarly, the Casimir energy of $N=3$ plates  with reflection coefficients $r_i,r_j,r_k=r_{D''(N'')}$ (Eqs. (36), (37)) from Eq. (21) is \begin{multline}
\frac{\Delta E_{(ijk)}}{A} = \frac{1}{2} \int_{-\infty}^\infty \frac{d\zeta}{2\pi}
\int \frac{d^2k_\perp}{(2\pi)^2} \Bigg[ \ln \Big[(1 - r^H_ir^H_j e^{-2 \kappa a})(1 - r^H_jr^H_k e^{-2 \kappa a})-(r^H_i (t^H_{j})^2 r^H_k) e^{-4 \kappa a})\Big] \\+ \ln \Big[(1 - r^E_ir^E_j e^{-2 \kappa a})(1 - r^E_jr^E_k e^{-2 \kappa a})-(r^E_i (t^E_{j})^2 r^E_k) e^{-4 \kappa a})\Big]\Bigg].
\end{multline}
Expanding inside the logarithm, introducing the spherical polar coordinates and evaluation of $s$ integral gives \begin{multline}
 \frac{\Delta E_{(ijk)}}{|\Delta E_{(D'D')}|} =  -\frac{45}{\pi^4} 
\int_{0}^{1} dt  \Bigg[ \Big[\text{Li}_4\left(\frac{2 b_1}{\sqrt{a_1^2-4 b_1}-a_1}\right)+\text{Li}_4\left(-\frac{2 b_1}{\sqrt{a_1^2-4 b_1}-a_1}\right)\Big] \\+ \Big[\text{Li}_4\left(\frac{2 b_2}{\sqrt{a_2^2-4 b_2}-a_2}\right)+\text{Li}_4\left(-\frac{2 b_2}{\sqrt{a_2^2-4 b_2}-a_2}\right)\Big] \Bigg], 
\end{multline}
where \begin{equation}
    a_1=-r^H_j (r^H_i+r^H_k),\  a_2=-r^E_j (r^E_i+r^E_k),
\end{equation}
are coefficients of $e^{- s}$ inside the logarithm and
\begin{equation}
\ b_1=r^H_i r^H_k \left((r^H_j)^2-(t^H_j)^2\right),\ b_2=r^E_i r^E_k \left((r^E_j)^2-(t^E_j)^2\right),
\end{equation}
are coefficients of $e^{-2s}$ inside the logarithm. Further, we can numerically evaluate the $t$ integral for varying $\sigma$ values.

In this manner, we can numerically evaluate the Casimir energy of arbitrarily N $\delta$-plates with constant and isotropic conductivity $\sigma$ for electric and magnetic properties of plates as defined in Eq. (35).
 \section*{Acknowledgements}
We sincerely thank K. V. Shajesh and Prachi Parashar for discussing their work on $\delta$-function plates and for their ongoing assistance.
% --------
% --------
\bibliographystyle{ieeetr}
\bibliography{sample.bib}
\end{document}